\documentstyle[12pt,epsfig]{article}
\begin{document}
\setlength{\baselineskip}{0.30in}
\newcommand{\beq}{\begin{equation}}
\newcommand{\eeq}{\end{equation}}
\newcommand{\be}{\begin{equation}}
\newcommand{\ee}{\end{equation}}
\newcommand{\nt}{\nu_\tau}
\newcommand{\nnt}{n_{\nu_\tau}}
\newcommand{\rnt}{\rho_{\nu_\tau}}
\newcommand{\mnt}{m_{\nu_\tau}}
\newcommand{\tnt}{\tau_{\nu_\tau}}
\newcommand{\bi}{\bibitem}

{\hbox to\hsize{December, 1997 \hfill TAC-1997-034}
\begin{center}
\vglue .06in
{\Large \bf {Impact of massive tau-neutrinos on primordial nucleosynthesis. 
Exact calculations}}
\bigskip
\\{\bf A.D. Dolgov
\footnote{Also: ITEP, Bol. Cheremushkinskaya 25, Moscow 113259, Russia.}
\footnote{e-mail: dolgov@tac.dk}, 
S.H. Hansen\footnote{e-mail: sthansen@tac.dk}
 \\[.05in]
{\it{Teoretisk Astrofysik Center\\
 Juliane Maries Vej 30, DK-2100, Copenhagen, Denmark}
}}
\\{\bf D.V. Semikoz \footnote{e-mail: semikoz@ms2.inr.ac.ru}} \\
{\it{Institute of Nuclear Research of the Russian Academy of Sciences\\
 60th October Anniversary Prospect 7a , Moscow 117312, Russia}}
\\[.40in]

\end{center}
\begin{abstract}
The influence of a massive Majorana $\nt$ on primordial nucleosynthesis
is rigorously calculated. The system of three integro-differential kinetic 
equations is solved numerically for $\mnt$ in the interval from 0 to 20 MeV. 
It is found that the usual assumption of kinetic equilibrium is strongly 
violated and non-equilibrium corrections considerably amplify the effect. Even 
a very weak restriction from nucleosynthesis, allowing for one extra 
massless neutrino 
species, permits to conclude that $m_{\nu_\tau} < 1 $ MeV. For a stricter
bound, e.g. for $\Delta N_\nu < 0.3$, the limit is $m_{\nu_\tau} < 0.35 $ 
MeV.\\

\end{abstract}
\bigskip

Pacs: 13.15.+g, 14.60.Pq, 26.35.+c 


\newpage
\section{Introduction}

The mass of the tau-neutrino is very loosely bound by 
experiments \cite{mnutau}:
\be
m_{\nu_\tau} < 18 \, {\rm MeV},
\label{mnut}
\ee
whereas it is much stronger restricted by cosmology. 
In the case that the $\nu_\tau$ is stable
on the scale of the universe age, $t_U \approx 10 $ Gyr, its mass obeys
the Gerstein-Zel'dovich limit~\cite{gz}, i.e. roughly $\mnt  < 10$ eV.  
If it is unstable, the cosmological limit is much less restrictive and depends
upon the  life-time of $\nt$ and  possible decay channels. 

If the $\nt$ is unstable on the cosmological time scale, so that
the Gerstein-Zel'dovich limit is avoided, but is stable on the 
nucleosynthesis time scale, i.e. $\tnt > 200$ sec, then considerations of 
primordial nucleosynthesis lead to a much better bound than (\ref{mnut}).
The bound on the $\nt$-mass from nucleosynthesis was first 
found in ref.~\cite{ktcs} and in a slightly
improved form in ref.~\cite{dr}. The calculations of the second work predict
a somewhat larger value of the frozen energy density of $\nt$, but in the 
translation of this result to the effective number of neutrino species, found
from the distortion of $^4He$ abundance, a numerical error was done which 
resulted in an overestimation of the number of additional effective neutrino
species. Still, even with the correction of this error the results of 
ref.~\cite{dr} are stronger than those of the pioneering paper~\cite{ktcs}. 
The calculations of both papers were 
done under the following basic assumptions.
It was assumed that  the massive $\nt$ and  the two
massless neutrinos, $\nu_e$ and $\nu_\mu$, are in complete 
kinetic equilibrium so that their energy distributions are
given by the canonical expressions:
\beq{
f_j(E)= {1 \over \exp (\beta E - \xi_j) + 1 }\, ,
\label{fed}
}\eeq 
where $\beta =1/T$ is the inverse temperature and $\xi_j$ are dimensionless 
effective chemical potentials. Another two simplifying assumptions were made, 
namely that the chemical potentials of  massless neutrinos are zero, 
$\xi_{\nu_e} = \xi_{\nu_\mu}=0$, and that the distribution function of
$\nt$ can be 
approximated by its Boltzmann limit:
\beq{
f(E)= \exp[(\mu(t) - E) /T(t)] \ll 1,
\label{feb}
}\eeq 
which is accurate when the temperature is small in comparison with the 
mass, $m>T$. Here we use the standard notation for the chemical potential,
$\mu \equiv \xi T$. 

It is well known that with these approximations the system of complicated 
integro-differential kinetic equations (see eq.~(\ref{dtf1}) below)
is reduced to a single ordinary differential equation (eq.~(\ref{dotnnt}))
for the 
number density of $\nt$ or, equivalently, for the unknown function $\xi (t)$ 
(see e.g. books \cite{zn,kt}), which  easily can be solved numerically. 
A similar approach was  recently used in ref.~\cite{rrw} where the role 
of a massive $\nt$ in the production of all primordial light elements and not 
only of $^4He$, as in refs. \cite{ktcs,dr}, was considered. 

The assumption of vanishing  $\xi_{\nu_e}$ and $\xi_{\nu_\mu}$ was relaxed
in ref.~\cite{fko}. 
If in addition one assumes validity of Boltzmann
statistics, then instead of one differential equation for $\xi_{\nu_e}$
there appear three ordinary coupled equations for three chemical potentials
$\xi_{j}(t), j=e, \mu, \tau$.  If one uses 
exact  Fermi-Dirac statistics, then  the unknown functions $\xi_{j}(t)$
remain ``inside'' the collision integrals but the kinetic equations still
remain ordinary differential equations in this
approximation, and not 
integro-differential ones for $f_j(t,p)$ as in the exact case. This
approach permits one 
to take into account an {\it average}
heating of massless neutrinos by annihilation of  
massive tau-neutrinos. It naturally
results in a smaller neutron-to-proton ratio and in a weaker influence 
of  a possible non-zero $\mnt$ on primordial nucleosynthesis. 
On the other hand, one can easily see that nonzero $\xi_{\nu_e}$ and
$\xi_{\nu_\mu}$ (in kinetic equilibrium) give rise to a considerably
larger frozen number density of $\nt$. 
Correspondingly the net effect on nucleosynthesis is stronger. 
We do indeed obtain a large increase in $n_{\nu_\tau}$, however,
a $\chi^2$-fit, for different $\nt$ masses, of our 
resulting distribution functions 
for the massless neutrinos to exact FD-distributions allowing for both an 
effective temperature
(possibly different from the expected neutrino 
temperature) and a chemical potential gives $\xi_{\nu_e}$
in the range from $+10^{-2}$ to $-2 \cdot 10^{-2}$ with extremum 
near $\mnt = 5 MeV$. 

It was found in ref.~\cite{dpv} that together with an overall neutrino 
heating there are considerable distortions at the high energy tail of  
massless neutrino spectra, created by annihilation of heavy tau-neutrinos
into massless (or light) $\nu$'s. The spectral distortion is
especially important for the electronic
neutrinos because they directly influence the frozen $n/p$-ratio, 
through the reactions: 
\be
\nu_e n \leftrightarrow pe^- \,\,\, {\rm and} \,\,\,
e^+ n \leftrightarrow p \bar \nu_e \,. 
\label{npre}
\ee
The calculations of the spectral distortions
were made in the Boltzmann limit and the problem was reduced
to the solution of an ordinary differential equation. 
According to ref.~\cite{dpv} the deviation from kinetic equilibrium
of the electronic neutrinos
is quite essential and has a noticeable 
influence on nucleosynthesis. 
These calculations were refined in~\cite{kk}.
The calculations of the present paper are
in agreement with the semi-analytical estimates of the distortion of 
$\nu_e$ spectrum made in refs.~\cite{dpv,kk}, which results in a 
larger $n/p$-ratio. The spectral distortion is not just a peak 
from $\nt$-annihilation, since the excessive energy
is redistributed by elastic scattering over the whole spectra 
of $\nu_{e,\mu}$. This effect gives rise to an overall neutrino heating 
or, in other words, to larger temperatures of the massless neutrinos relative
to the photon temperature.  This leads in turn to a later freezing of the 
reactions (\ref{npre}) and to a lower $n/p$-ratio. According to 
our calculations the average heating produces a stronger effect than the 
spectral distortion. This agrees with the statement made in 
ref.~\cite{kk} (see also "Note added" to ref.~\cite{fko}).

In view of this discussion it is interesting to solve the exact system of
integro-differential kinetic equations numerically (with a good precision) 
and to find exactly, without any simplifying assumptions,
the impact of possibly massive tau-neutrinos 
on primordial nucleosynthesis. The first calculations of this kind were done 
in ref.~\cite{hm}. In what
follows we will correct some matrix elements for scattering of Majorana 
neutrinos presented in ref.~\cite{hm}, and solve  
the kinetic equations numerically, using a more accurate code. 
Our results for $n_{\nt}$ are in a good agreement with ref.~\cite{hm}. We also 
agree with this paper in the calculation of the changes in light element 
abundances for low values of $\mnt$. In the high mass range our results
are noticeably lower. We ascribe this difference to a lower momentum
cut-off made in ref.~\cite{hm}, $y_{max} \approx (p/T)_{max} \approx 13$,
in the region being 
essential for the reactions (\ref{npre}), while we made the 
calculations with a considerably larger cut-off, $y_{max}=20$. 
 
Our technique repeats that of ref.~\cite{dhs},
where similar calculations were done for massless neutrinos,
with evident complications in the case of a massive $\nt$ due to a larger 
number of equations and different matrix elements. We conclude that the 
influence of non-equilibrium corrections on primordial 
nucleosynthesis is considerably larger than what was calculated in the 
earlier papers under assumptions of kinetic equilibrium and Boltzmann
statistics. The dominant effect comes from an increase of the frozen energy 
density of massive $\nt$.

\section{Qualitative discussion}

The complete set of integro-differential kinetic equations has the form:
\beq{
 (\partial_t - Hp\partial_p) f_j (p_j,\,t) = I^{coll}_{j},
\label{dtf1}
}\eeq
where the collision integral for two-body reactions
$1+2 \rightarrow 3+4$ is given by the expression:
\begin{eqnarray}
I^{coll}_{1} = {1\over 2E_1}\sum \int {d^3 p_2 \over 2E_2 (2\pi)^3}
{d^3 p_3 \over 2E_3 (2\pi)^3}{d^3 p_4 \over 2E_4 (2\pi)^3}\,
S\, |A|^2_{12\rightarrow 34}
\nonumber \\
(2\pi)^4\delta^{(4)} (p_1+p_2-p_3-p_4) F(f_1,f_2,f_3,f_4)\, ,
\label{icoll}
\end{eqnarray}
where $F = f_3 f_4 (1-f_1)(1-f_2)-f_1 f_2 (1-f_3)(1-f_4)$,
$|A|^2$ is the weak interaction amplitude squared summed over spins of
all particles, and $S$
is the symmetrization factor which includes 
$1/2$ from the averaging over the 
first particle, $1/2!$ for each pair of identical
particles in the 
initial and final states and the factor 2 if there are 2 identical
particles in the initial state; the summation is done over all possible
sets of leptons 2, 3, and 4.

Numerical solution of such a set of equations is tremendously more difficult 
than the solution of the well known simple ordinary differential equation 
for the 
number density of massive neutrinos, which is valid in the case when the
neutrinos are in kinetic equilibrium and obey Boltzmann statistics:
\beq
\dot \nnt = \langle \sigma_{ann} v \rangle ( n^2_{eq} - \nnt^2 ).
\label{dotnnt}
\eeq
One would naturally ask if such a complexity is necessary. Eq.~(\ref{dotnnt})
gives a very good approximation if the rate of elastic scattering of the
massive particles in question is much larger than their annihilation rate.
This is usually true  for very non-relativistic particles since their 
annihilation is Boltzmann suppressed while elastic scattering is not. For
the case of $\nt$, both elastic scattering and annihilation are frozen at
temperatures which are not much below $\mnt$ and rather close to each other
so the effects of deviations from equilibrium may be 
significant. When one calculates the frozen number density of massive $\nt$,
a gradual switch-off of elastic scattering results in a distortion of their
spectrum and in a faster-than-equilibrium cooling.
Since the cross-section of their annihilation is proportional
to the  energy squared, this faster-than-equilibrium cooling gives rise to
an increase in their frozen number and energy densities and results 
in a larger effect 
on nucleosynthesis. A simple attempt to take this non-equilibrium cooling 
into account was made in ref.~\cite{dr} where it was assumed that at 
temperatures above a certain value, $T_{eq}$, which is determined by the
strength of the elastic scattering, 
tau-neutrinos have the equilibrium distribution
in energy, $f_{\nt} = \exp [-\sqrt{p^2 +\mnt ^2}/T + \xi ]$,
while below $T=T_{eq}$ the complete kinetic decoupling was assumed,
so that the momentum $p$ was scaled as inverse universe expansion 
factor $a(t)$.
Unfortunately this simple anzats is not very accurate and more elaborate 
calculations (ref.~\cite{ad}) 
show, that the spectrum is distorted in a more complicated
way. Rough estimates of paper~\cite{ad} show that these non-equilibrium 
corrections give rise to an increase of $\nnt$ at the level of 10\%. 
As we see below in fig.~1,  the calculations of the present paper show
an even bigger effect. The frozen number density of heavy $\nt$
differ from that of ref.~\cite{ktcs} by $35$ \% and from that of 
ref.~\cite{dr} by $15$ \% in the maximum.

Another important effect is a distortion of the equilibrium spectra of
massless neutrinos and in particular of electronic neutrinos. An
excess of $\nu_e$ at the high energy tail of the spectrum gives rise
to a larger frozen number density of neutrons 
and correspondingly to a larger abundance of
$^4He$.  There is also an opposite effect related to an overall
increase in the number density of 
electronic neutrinos due to the annihilation $\nt \nt
\rightarrow \nu_e \nu_e$. This results in a later neutron decoupling
and to a smaller frozen ratio $n/p =\exp (-\Delta m /T_f)$, where
$T_f$ is the freezing temperature of the ($n\leftrightarrow
p$)-reactions and $\Delta m= 1.293$ MeV is the neutron-proton mass
difference. It is difficult to separate rigorously these two effects
using the $\nu_e$ spectrum calculated here, because of an ambiguity in
the choice of the unperturbed distribution, in particular because the
effective temperature is different in different parts of the spectrum.

It is also important that the ratios of the photon temperature $T_\gamma$ and
the (average) temperatures of the massless neutrinos are different 
from the canonical value $T_\gamma /T_\nu = (11/4)^{1/3}$. It can easily be 
understood that these ratios depend non-monotonically  on $\mnt$. 
They go up with $\mnt$, reach maximum near $\mnt = 5$ MeV
and fall down close to the massless value for large $\mnt$.

One sees that there are several essential effects associated with the 
distortion
of kinetic equilibrium which significantly may change the primordial abundances
of $^4He$ and other light elements. All of them are automatically taken into 
account in our calculations below.

\section{Kinetic equations}

We want to solve kinetic equations numerically  
for the following system of interacting particles:  
massless $\nu_e$ and $\nu_\mu$, massive $\nt$, electron-positron pairs, and
quanta of electro-magnetic radiation. There are three unknown distribution
function of time and momenta, $f_j (t,p) $, for the three neutrino species.
We assume that electron-positrons and photons are in complete equilibrium
so that their distributions are given by eq.~(\ref{fed}) with vanishing 
chemical  potentials and with temperature $T(t)$, which is an unknown 
function of time (of course photons obey Bose statistics).
The assumption of equilibrium of   the
electro-magnetic component of the primeval plasma
is very accurate because of a large strength of the
electro-magnetic interactions.
Thus there are 3 unknown functions of $t$ and $p$ and one unknown function of
$t$. For their determination we have three kinetic equations (\ref{dtf1}).  
The fourth necessary equation is the covariant conservation of energy-momentum
in the expanding universe:
\beq
\dot \rho = -3H (\rho +P) \, .
\label{dotrho}
\eeq
Usually one determines the temporal evolution of temperature from the law
of entropy conservation. However, the latter is only true  in thermal
equilibrium, so if one wants to describe deviations from equilibrium
consistently, a more general, though more complicated eq.~(\ref{dotrho}) 
must be used. In this equation 
$\rho$ and $P$ are respectively total energy and pressure densities in the
cosmic plasma. They are given by the expressions:
\beq{
 \rho = {\pi^2 T^4_\gamma\over 15}  + \int {2dq q^2
\sqrt{q^2 + m^2_e} \over \pi^2 [\exp {(E/T_\gamma)} +1] } +
  \sum_{j=\nu_e,\nu_\mu }\int {dq q^3 f_j \over \pi^2}
+ \int {dq q^2\over \pi^2} \sqrt{q^2 +\mnt^2}\, f_{\nu_\tau},
\label{rho}
}\eeq
and:
\beq{
 P = {\pi^2 T^4_\gamma\over 45}  +   \int {2dq q^4  
{(q^2 + m^2_e)}^{-1/2} \over 3\pi^2 [\exp (E/T_\gamma) +1 ]} +
\sum_{j=\nu_e,\nu_\mu }\int {dq \,q^3 f_j\over 3\pi^2} +
\int {dq\, q^4 f_{\nt}  \over 3\pi^2 \sqrt{ q^2 + \mnt^2}}.
\label{p}
}\eeq
The Hubble parameter, $H=\dot a /a$, is related to the total energy
density $\rho$ in the usual way:
\beq
H^2 = {8\pi \rho \over 3 m_{Pl}^2},
\label{h2}
\eeq
where $a(t)$ is the universe expansion factor (scale factor) and 
$m_{Pl} = 1.22 \cdot 10^{19}$ GeV is the Planck mass.
 
It is convenient  instead of time and momenta to use
the following dimensionless
variables:
\beq{
 x= m a(t), \,\,  y_j= p_j a(t),
\label{xy}
}\eeq
where $m$ is an arbitrary parameter with dimension of mass,
which we took as $m=1$~MeV, and the
scale factor $a(t)$ is normalized so that $a(t) = 1/T_\nu = 1/T_\gamma$ at high
temperatures or at early times. In terms of these variables the kinetic
equations (\ref{dtf1}) can be rewritten as:
\beq{
 Hx \partial_x f_j(x,y_1) = I^{coll}_j.
\label{hxdxf}
}\eeq
The relevant reactions and the corresponding matrix elements squared are
presented in tables 1 and 2 for the cases when the first particle is
$\nu_e$ ($\nu_\mu$) and $\nt$ respectively. There are the following differences
with ref.~\cite{hm}: 1) for the reactions $\nu_a\nu_a\rightarrow \nu_a \nu_a$ 
we took twice larger contribution because of identical particles in the 
initial state; 2)~we have a different expression for the matrix element squared
for elastic scattering of massive Majorana $\nt$.

Our procedure of solving these equations is essentially the same as in our
previous paper \cite{dhs}, where non-equilibrium corrections to the 
spectra were
calculated for the case when all neutrinos were 
massless. The collision integral
is reduced from 9 down to 2 dimensions by the method described in
ref.~\cite{st} with some complications connected with the momentum
dependence of the matrix elements (see ref.~\cite{dhs}). The results of the 
calculations are presented in Appendix A. 

\subsection{The problem of initial conditions}

We solve the system of kinetic equations (\ref{hxdxf}) in the ``time'' 
interval  $x_{in} \le x \le x_{out}$.
All the collision integrals in the r.h.s. of  eq.~(\ref{hxdxf}) 
are suppressed at large $x$ at least by the factor $1/x^2$.
We find that at $x \approx 50$  all the variables, we are 
interested in, reach their asymptotical values so, to be on the safe
side, we choose the final time $x_{out} = 100$. 

The nucleosynthesis code of ref.~\cite{kaw} requires 
the final time $x \approx 2000$ for all
energy densities and $n \leftrightarrow p$ rates. We applied 
a separate program,
which calculates these quantities, using the energy conservation law and
 our final values for the distribution 
functions $f_{\nu_i}$ and the temperature $T_\gamma$. 

At early times, $x \ll 1$, the collision rates  in  eq.~(\ref{hxdxf}) 
are very high due to the factor $1/x^4$  and therefore all  
non-equilibrium corrections to the distribution functions are suppressed. 
Correspondingly we could start at some initial "moment" 
$x_{in} \ll 1$ with the equilibrium distribution functions:
\be
f_{\nu_e (\nu_\mu)} = \frac{1}{e^{p/T_{\nu_e (\nu_\mu)}}+1}~, ~~~~~
f_{\nu_\tau} = \frac{1}{e^{(E_{\nu_\tau}- \mu)/T_{\nu_\tau}}+1}~, ~~~~~ 
f_{e^{\pm}} = \frac{1}{e^{E_e/T_\gamma}+1}~, 
\label{incond}
\ee   
where $T_j$ are the temperatures of the 
particles, $\mu$ is the chemical potential of 
$\nt$ and $E_j$ are the energies, $E_j = \sqrt{p^2 + m_j^2}$.

At very early times, when $T_\gamma \gg \mnt$, 
we can neglect the mass of $\nt$ 
and put $T_{\nu_e}=T_{\nu_\mu}=T_{\nu_\tau}=1/a = T_{\gamma}$ and $\mu = 0$.  
However, in the case of a large $\nt$ mass
(e.g. $\mnt = 20 MeV$) we should start at least at 
$x_{in} = 0.01$, in order 
to satisfy 
the condition $T_\gamma \gg \mnt$. Since the collision rates are very 
high at this time, the time steps in the 
numerical integration become very small
and the program  requires much CPU time.\footnote{CPU time, which is 
proportional to the number of time steps, grows at least
as fast as the rates, i.e. as $1/x_{in}^4$; for example, 
a run with $x_{in}=0.04$
requires twice as much CPU time as a run with $x_{in}=0.05$. }  
Another problem is that a big number of  small time steps leads to an
accumulation of numerical errors.

Thus it is very difficult to start at $T_\gamma \gg \mnt$, and we need to 
find reasonable and convenient
initial values of $T_j$ and $\mu$. We assume that at 
$T_\gamma \geq max(10 MeV, \mnt)$  
the collision rates  are high enough to keep all the neutrinos in 
kinetic equilibrium with the distributions given by eqs.~(\ref{incond}). 

A simple  self-consistent way to find the evolution of $T_j$ and $\mu$ in
the ``almost equilibrium'' region
is to integrate all the kinetic equations over external momenta and to
get the system of differential equations for the number densities $n_{\nu_j}$. 
However, we can express the collision integrals 
in terms of $n_{\nu_j}$ only in the limit of Boltzmann statistics, which 
differs from the exact Fermi-Dirac one by approximately 10\% \cite{dk}. 
Therefore in order to keep
a high precision in the calculations we need another approximation 
for the initial conditions. We require that the temperatures of all neutrino
species and the photon temperature are equal, $T_{\nu_j} = T_\gamma$,
and assume that the initial value of the chemical potential $\xi_{\nu_\tau}$
is zero.
The unknown value of the temperature  can be found from the energy 
conservation law (i.e. we do not require that  $T_\gamma = 1/a$).
Thus our initial conditions at $x=x_{in}$ are the following:
\be
f_{\nu_e (\nu_\mu)} = \frac{1}{e^{p/T_\gamma}+1}~, ~~~~~
f_{\nu_\tau} = \frac{1}{e^{E_{\nu_\tau}/T_\gamma}+1}~, ~~~~~ 
f_{e^{\pm}} = \frac{1}{e^{E_e/T_\gamma}+1}~, 
\label{initial}
\ee   
where $T_\gamma(x_{in}, \mnt)$ is found from the equation 
$\dot \rho = -3H (\rho +P)$, evolved from higher temperatures (or smaller
$x$), where expressions (\ref{initial}) are very accurate with ``preinitial'' 
$T_\gamma =1/a$.
Note, that for $\mnt > 1 MeV $ the photon temperature  $T_\gamma(x_{in}, \mnt)$
strongly depends on $\mnt$. 
     
In order to check the validity of our approximation we compared the energy 
densities and number densities for all neutrino species using distributions
given by eqs.~(\ref{initial}) with the same quantities found from the solution 
of the kinetic 
equations (\ref{hxdxf}) for the case $\mnt = 20 MeV $ in the time interval 
$0.04 < x < 0.05$. We found that both the energy 
densities and number densities differ in these two approaches 
only by $0.1 \% $.
 
For such values of neutrino mass where 
$\mnt > T_\gamma(x_{in}, \mnt)$ at $x_{in} = 0.1$,
non-equilibrium corrections cannot be neglected, so for $\mnt \ge 6 MeV$ 
we choose $x_{in}=0.05$, and for  smaller 
masses, $\mnt < 6 MeV$, we take $x_{in}=0.1$.    

A careful choice of initial values of 
$x_{in}$ and $T_\gamma(x_{in}, \mnt)$ is very important for the numerical
results. For example, if for $\mnt = 20 MeV$ we  take $x_{in}=0.1$ and 
choose to use the ``wrong'' temperature evolution 
$T_\gamma = 1/a$, then we  find that 
the function $ rm = \mnt \,  n_{\nu_\tau}/n_{eq} $  
is smaller by about 6\% than $rm$
calculated  with $x_{in}=0.05$. With our procedure using 
$T_\gamma \neq 1/a$, we are certain that the final results, in particular
the abundances of light elements, have 
reached a plateau as long as we use $x_{in} \leq 0.1$.
\subsection{Numerical integration of kinetic equations}

For the dimensionless momentum $y$ we took $100$ point grid equally 
spaced in the
region $0 \le y \le 20$. At $y=0$ the analytical expressions for the 
collision integrals differ from those at $y\neq 0$ due to 
$0/0$-uncertainty arising from the factor $1/E_1 p_1$ in front of the integrals
and similar vanishing factors in D-functions. Because of that we include the
point $y=0$ separately. This permits to compare the 
collision integrals at $y=0$ 
with those in nearby points, $y \ll 1$, in order to
check that numerical errors  are small in the region of small momentum,
$y<1$. This is especially important for a massive  
tau-neutrino, because its distribution function rapidly changes 
in the course of evolution.  

Evolution in time is calculated by the simple Euler method for small
$x$ ($x<1$)
and by the Bulirsch-Stoer method (see e.g. ref.~\cite{numrec}) 
for large $x$ ($x>1$). We checked
that numerical errors in the distribution functions calculated by
the Euler method for $x<1$ are smaller than $0.1\%$. The Euler method
allows us to save a factor of 2 in CPU time compared 
to the second order Runge-Kutta method
and more than a factor of 10  compared to the Bulirsch-Stoer method. 
For $x>1$ the situation is opposite. The Euler method requires very small 
time steps in order to be precise enough, while the more powerful 
Bulirsch-Stoer method requires fewer time steps in order to reach the 
same precision.   

An important check of our program is a run with a small value of
$\nt$ mass ($\mnt=0.1 MeV$) which is in a very close agreement with the 
results of our calculations in the massless case (see ref.~\cite{dhs}).

\section{Discussion of results}

The results of our calculations are presented in several figures below.
The energy density of massive $\nt$ is characterized by the quantity 
$r\mnt$, where $r= \nnt /n_{\nu_0}$ and $n_{\nu_0}$ is the number density of 
massless equilibrium neutrinos. This quantity is presented in fig.~1
for asymptotically large $x$. Our results are rather close to 
the previous calculations of refs.~\cite{ktcs,dr,hm} for small masses,
while for large masses they are considerably larger than those obtained
in refs.~\cite{ktcs,dr} and in fine agreement with the calculations of 
ref.~\cite{hm}. 
The energy density of $\nt$ influences the rate of
the universe cooling  and  affects two quantities which are
important for nucleosynthesis: the frozen neutron-to-proton
ratio and the time moment corresponding to $T \simeq 0.065$ MeV, 
when  formation of light elements begins. These
two quantities determine the neutron number density at the onset of
nucleosynthesis and through that
the abundances of light elements and in particular of
$^4He$. A larger $r$ results in a larger mass fraction of $^4He$ both because
of a faster cooling rate and because of a more efficient production of 
non-equilibrium electronic neutrinos \cite{dpv}.  
Of course the relative abundance of $\nt$ is a function of "time" $x$,
so in our numerical calculations we used the exact function $r(x)$ and not
just its asymptotic value. The same is true for the energy density, for 
which the exact expression (\ref{rho}) was used.

In fig. 2 the product $T_\gamma a$ is presented as a function of $x$
for different values of $\mnt$.
In the standard calculations this quantity  coincides with the ratio
$T_\gamma / T_\nu $, which tends to $1.40102$ for high $x$. 
For massless $\nt$ the product $T_\gamma a$ is slightly smaller than 
this result because of the 
energy transfer from the electro-magnetic components of the primeval plasma to 
the massless neutrinos. For a sufficiently heavy $\nt$ the product 
$T_\gamma a$ is larger than the canonical value because now the 
effect of photon heating by annihilation of
massive $\nt$ is stronger than the energy loss mentioned above. 
In this case massless neutrinos are heated more  than 
electrons, 
because electrons have to share the given energy with  photons.
For very heavy $\nt$'s the
final photon temperature ($T_\gamma \; a$) can be estimated from entropy
conservation to be $(43/36)^{1/3}$ times bigger
than in the massless case.

In fig.~3 we plot the  relative deviations of the energy densities of
$\nu_e$ and $\nu_\mu$ from the equilibrium value,
$\delta \rho/\rho = (\rho - \rho_{eq})/\rho_{eq}$, 
where $\rho_{eq} (x) = (7/120 \pi^2)(1 MeV/x)^4 $. 
The solid and dashed lines in fig.~3 correspond to $\nu_e$ and $\nu_\mu$
respectively.
In fig.~3a  the evolution  of $\delta \rho/\rho$ as functions
of   $x$ for several different masses of $\nu_\tau$ is presented.
In fig.~3b the asymptotic
values of  $\delta \rho/\rho$ as functions of $\mnt$ are shown. 
An explanation of the different and non-monotonic relative
behavior of $\delta \rho_{\nu_e}/\rho$ and 
$\delta \rho_{\nu_\mu}/\rho$ is given below.

It is convenient to characterize a distortion of neutrino spectrum by the
average neutrino energy, $\langle E \rangle = \rho / n$.
In the equilibrium case it is equal to $\langle E_{eq}\rangle  =3.15 T$. 
In fig. 4 we present the relative deviation 
$(\langle E \rangle -\langle E_{eq}\rangle )/\langle E_{eq}\rangle $ for 
electronic and muonic neutrinos. For small masses of $\nt$
this quantity is practically mass independent. In this mass range  
$\nu_e$ and $\nu_\mu$  are heated by $e^-e^+$-annihilation. 
Electronic neutrinos are heated more efficiently because of a larger coupling
to electrons/positrons due to charged current interactions. For higher $\mnt$
the situation is opposite. As we have already mentioned, with a heavy $\nt$,
massless neutrinos ($\nu_e$ and $\nu_\mu$)
are "overheated", so that their excessive energy is 
transferred to the electro-magnetic component of the primeval plasma
($e^-\,,e^+\,,\gamma$). Now a
stronger coupling of $\nu_e$ to electrons results in a more efficient 
cooling of $\nu_e$ than $\nu_\mu$. This explains why muonic neutrinos are  
heated less than $\nu_e$ for a low $\mnt$, while for a higher $\mnt$ the
situation is reversed. However, for a very large $\mnt$ we return to 
hotter $\nu _e$. This is because  very heavy tau neutrinos
effectively stopped annihilating while $\nu_e, \nu_\mu$ and $e^\pm$ 
were still in equilibrium. At that stage
their temperatures were the same and energy was not transferred either way.
At smaller temperatures when $\nu_e$ and $\nu_\mu$ were already almost
decoupled, the ``old'' process of
$e^+ e^-$ annihilation heated up $\nu_e$ more than $\nu_\mu$.

The shape of the 
electronic neutrino spectrum, disturbed by $\nt$ annihilation,
is presented in figs.~5 and 6. 
For large masses the relative distortion 
$\delta$ is much larger than 1 for high momentum, 
so we present the distorted function itself 
and compare it with the equilibrium one (fig.~5). 
The magnitude and in particular the non-monotonic behavior of the spectral 
distortions as functions of $x$ for different masses ($m=7$ MeV and $m=20$ MeV)
are in agreement with the semi-analytical calculations of refs.~\cite{dpv,kk}.
The effect of the changed neutrino distribution on light element abundance 
is essentially $y^2 \delta f$, which is plotted as a function of y for
different neutrino masses in fig.~6.

In order to extract  implications of a massive $\nt$ on light
element abundance, we have modified the standard nucleosynthesis code 
(ref.~\cite{kaw})
in the following way. We calculate various quantities at the 
correct photon temperature and import the values to the code.
These imported quantities are the neutrino energy densities $\rho_{\nu_i}$,
the 6 weak interaction rates for $(n \leftrightarrow p)$-reactions
(this includes reactions (\ref{npre}) and decay and inverse decay),
and finally $d (\ln \, a^3) /d T_\gamma$ (with the account of neutrinos) 
which governs the evolution of the photon temperature.

The effect of a nonzero $\mnt$
on the abundances of light elements is illustrated  
in figs.~7 and 8. Comparing our results with 
the standard calculations of e.g. helium production
with a variable number of massless neutrino species,
we express the impact of massive $\nt$ on nucleosynthesis
in terms of the number of equivalent
neutrino species $\Delta N = N_{eff} -3$~\cite{nuclcode}. 
The excessive number of neutrino
species found from 
$^4 He$ is  presented in fig.~7, where our  results are compared to those of 
the previous papers~\cite{ktcs,hm,kimmopriv}. 
For low masses we are in a good agreement with ref.~\cite{hm},
while for high masses our results are noticeably lower. This difference is
probably explained by a lower momentum cut-off made in ref.~\cite{hm}.
The authors 
calculated spectral functions of different neutrinos up to 
$y_{max} = 12 T_\gamma /T_\nu$, while we were able to do that up to 
$y_{max} =20$. When we diminished our cut-off down to the value of
ref.~\cite{hm}, we reproduced their higher results in the high mass region.  
A large  sensitivity of $^4He$-production to the value of $y_{max}$
is connected to a strong (quadratic) energy dependence of the cross-sections
of $n\leftrightarrow p$ reactions. 
To independently test the effect of the distortion of the high energy 
neutrinos we calculated the neutron to proton ratio as a function of 
the cut-off of the distortion, numerically solving the kinetic equation 
for the $n/p$-ratio. The results is, that the $n/p$ ratio is decreasing  
with an increase of momentum cut-off. This continues until rather high 
momenta (around $y=16$ for $\mnt=20$) when the effect changes sign, so
that $n/p$ is increasing when we increase the cut-off in $y$ from 16 to 
20 (as predicted by ref.~\cite{dpv} and refined by ref.~\cite{kk}). 
The net result is, that the increase of the cut-off from $y=12$ to 20 
leads to a decrease of $n/p$ in agreement with the full nucleosynthesis 
calculation. With our cut-off in momentum, $y_{max} =20$, the error in 
$\Delta N_\nu$ is approximately an order of magnitude smaller than that 
at $y_{max} =12$, i.e. $\delta\,(\Delta N_\nu)< 0.02$.
 
In fig.~8  the number of effective extra neutrino species  
obtained from other light elements: $^2H,\, ^3He$ and $^7 Li$ is presented,
using $\eta_{10} = 3.0$. 
These curves are given for illustration only because the best bound on
$N_\nu$ is obtained from $^4 He$. The results are in a qualitative
agreement with the simplified calculations of ref.~\cite{rrw}, though there are
quantitative differences. 

The recent confusion with the data on abundance of primordial deuterium 
(ref.~\cite{deut})
makes it difficult to put a stringent bound on $\Delta N $. It seems that 
it is rather safe to conclude that $\Delta N <1$. 
In this case the consideration of primordial nucleosynthesis excludes the 
mass of $\nt$ in the interval $1- 22$ MeV. Recall that it is valid for the
sufficiently long-lived $\nt$, i.e. for $\tau_{\nt} > 200$ sec.
Together with the direct experimental
bound (\ref{mnut})  it
gives $\mnt < 1$ MeV. This result is obtained for $\eta _{10}=3.0$. At
lower $\eta _{10} = 1-2$ the lower bound is slightly strengthend.
Hopefully a resolution of the 
observational controversies in the light element abundances will permit to 
shift this limit to even smaller values of $\mnt$. 
The result in fig.~7 is exact in the sense, that it is independent of
any measured abundance. One can therefore choose ones favorite 
experiment, extract a $dN$-limit and apply it to fig.~7.
In particular, if 
the limit on $\Delta N_\nu$ would return to the "good old" value,
$\Delta N_\nu < 0.3$, one could conclude from fig.~7 that $\mnt < 0.35$ MeV.
If one believes in $\Delta N_\nu < 0.2$ (see e.g. ref.~\cite{nucl}) one
finds $\mnt < 0.25$ MeV, and in
the case of an over-optimistic limit, $\Delta N_\nu < 0.1$, which was 
advocated in the literature a few years ago, we obtain $\mnt < 0.15$ MeV.

\bigskip

{\bf Acknowledgment}\\
We wish to thank I.I. Tkachev for valuable comments
and S. Pastor for pointing out an error in an early version of the
program~\cite{dhs}. We also thank S. Hannestad and J. Madsen for discussions.
The work of AD and SH was supported by the Danish National Science 
Research Council through its
funding of the Theoretical Astrophysical Center.
The work of DS was supported in part 
by the Russian Foundation for Fundamental Research through grants 95-02-04911A
and 97-02-17064A.
DS thanks the Theoretical Astrophysical Center for hospitality during the last 
stages of this work.

\appendix

\section{Kinetic equations}
  
After reducing the collision integral for two-body reactions 
in eq.~(\ref{dtf1}) from nine to two dimensions the 3 kinetic
equations for the distribution functions are
solved numerically. For the sake of brevity we introduce some
notation: $f_a(p_j) \equiv f_a^{(j)}, d_1 = D_1, d_2(3,4) = 
D_2(3,4)/E_3 E_4$ and $d_3 = D_3/E_1 E_2 E_3 E_4$, with  
D-functions defined in Appendix A of ref.~\cite{dhs}. 
The functions $d_1$ and $d_3$ are symmetric in all four arguments. 
We introduce one more condensed notation:
\be
F(f^{(1)},f^{(2)},f^{(3)},f^{(4)}) = f^{(3)} f^{(4)}
(1 -f^{(1)})(1 -f^{(2)}) - f^{(1)} f^{(2)}(1 -f^{(3)})(1 -f^{(4)}),
\ee
and write the coupled system of kinetic equations  as follows:
\begin{eqnarray}
 Hx\partial_x f_{\nu_e} ^{(1)} = { G^2_F \over 2\pi^3 p_1} 
\int dp_2 p_2 dp_3 p_3 dp_4 p_4 \delta (E_1 +E_2 -E_3-E_4)
\nonumber \\
\left\{ F(f^{(1)}_{\nu_e},f^{(2)}_{\nu_e},f^{(3)}_{\nu_e},f^{(4)}_{\nu_e})
\right.
\nonumber \\
\left[ 6 d_1 + 2 d_2(1,2) + 2d_2(3,4) - 2d_2 (1,4) - 2d_2 (2,3)
+ 6 d_3 \right]
\nonumber \\
+ \left[ 
F(f^{(1)}_{\nu_e},f^{(2)}_{\nu_e},f^{(3)}_{\nu_\tau},f^{(4)}_{\nu_\tau})
+F(f^{(1)}_{\nu_e},f^{(2)}_{\nu_e},f^{(3)}_{\nu_\mu},f^{(4)}_{\nu_\mu}) \right]
\nonumber \\
\left[ d_1 - d_2(1,4)-d_2(2,3) + d_3 \right]
\nonumber \\
- F(f^{(1)}_{\nu_e},f^{(2)}_{\nu_e},f^{(3)}_{\nu_\tau},f^{(4)}_{\nu_\tau})
\left[ \mnt^2/2 (d_1 + d_2(1,2))/E_3E_4 \right]
\nonumber \\
+ \left[ F(f^{(1)}_{\nu_e},f^{(2)}_{\nu_\mu},f^{(3)}_{\nu_e},f^{(4)}_{\nu_\mu})
+F(f^{(1)}_{\nu_e},f^{(2)}_{\nu_\tau},f^{(3)}_{\nu_e},f^{(4)}_{\nu_\tau})
\right] \nonumber \\
\left[2 d_1 + d_2(1,2)+ d_2(3,4)- d_2(1,4)- d_2(2,3) + 2d_3
\right]
\nonumber \\
+F(f^{(1)}_{\nu_e},f^{(2)}_{\nu_\tau},f^{(3)}_{\nu_e},f^{(4)}_{\nu_\tau})
\left[ \mnt^2 (d_1 - d_2(1,3))/E_2E_4  \right]
\nonumber \\
+F(f^{(1)}_{\nu_e},f^{(2)}_{\nu_e},f^{(3)}_{e},f^{(4)}_{e})
\left[4(g_L^2 + g_R^2) \left( d_1 - d_2(1,4) - d_2(2,3)  
+ d_3  \right) \right.
\nonumber \\
+ 4 g_L g_R m_e^2 \left. \left( d_1 + d_2(1,2)\right) /E_3E_4 \right]
\nonumber \\
+F(f^{(1)}_{\nu_e},f^{(2)}_{e},f^{(3)}_{\nu_e},f^{(4)}_{e})
\left[ 4  (g_L^2 + g_R^2) \left(2 d_1 + d_2(1,2) + d_2(3,4) \right. \right.
\nonumber  \\
\left. - d_2(1,4)  - d_2(2,3) + d_3 \right)  
\left. \left. - 8   g_L g_R m_e^2 \left( d_1 - d_2(1,3)\right)/E_2E_4 \right]
\right\}. \nonumber
\end{eqnarray}
The kinetic equation for $f_{\nu_\mu}$ has the same form with the 
substitutions $f_{\nu_e} \leftrightarrow 
f_{\nu_{\mu}}$ and $g_{L} \rightarrow \tilde{g}_{L} =g_{L} - 1$.

The equation for $f_{\nu_\tau}$ reads:
\begin{eqnarray}
 Hx\partial_x f_{\nu_\tau} ^{(1)} = { G^2_F \over 2\pi^3 p_1} 
\int dp_2 p_2 dp_3 p_3 dp_4 p_4 \delta (E_1 +E_2 -E_3-E_4)
\nonumber \\
\left\{ F(f^{(1)}_{\nu_\tau},f^{(2)}_{\nu_\tau},f^{(3)}_{\nu_\tau},
f^{(4)}_{\nu_\tau})
\right.
\nonumber \\
\left[ 6 d_1  + 2 d_2(1,2) + 2d_2(3,4) - 2d_2 (1,4) - 2d_2 (2,3)
+ 6 d_3 \right. \nonumber \\
\left. + 6 d_1 \mnt^4 /E_1 E_2 E_3 E_4 + 4 \mnt^2 \left(
2(d_1 - d_2 (1,4))/E_2 E_3 - (d_1 + d_2(1,2))/E_3E_4
\right) \right] \nonumber \\
+ \left[ 
F(f^{(1)}_{\nu_\tau},f^{(2)}_{\nu_\tau},f^{(3)}_{\nu_e},f^{(4)}_{\nu_e})
+F(f^{(1)}_{\nu_\tau},f^{(2)}_{\nu_\tau},f^{(3)}_{\nu_\mu},f^{(4)}_{\nu_\mu}) \right]
\nonumber \\
\left[ d_1 -d_2(1,4)- d_2(2,3)
+ d_3  - \right. \nonumber \\
\left. - m _\tau^2 /2 \left(d_1  + d_2(3,4)/E_1E_2 \right)\right]
\nonumber \\
+F(f^{(1)}_{\nu_\tau},f^{(2)}_{\nu_\tau},f^{(3)}_{e},f^{(4)}_{e})
\left[4 (\tilde{g}_L^2 + g_R^2) \right. \nonumber \\
\left(d_1 - d_2(1,4) - d_2(2,3)    + d_3
- \mnt^2 /2(d_1 + d_2(3,4))/E_1E_2\right) 
\nonumber \\
+ 4 \tilde{g}_L g_R m_e^2 \left. \left( d_1 + d_2(1,2) 
- 2 \mnt^2 d_1/E_1E_2\right)/E_3E_4 \right]
\nonumber \\
+F(f^{(1)}_{\nu_\tau},f^{(2)}_{e},f^{(3)}_{\nu_\tau},f^{(4)}_{e})
\left[ 4  (\tilde{g}_L^2 + g_R^2) \right. \nonumber \\ 
\left(2 d_1 + d_2(1,2) + d_2(3,4)  
- d_2(1,4)  - d_2(2,3) + 2 d_3 + \mnt^2 (d_1 - d_2(2,4))/E_1E_3 
\right)  
\nonumber \\
\left. \left. - 8   \tilde{g}_L g_R m_e^2 \left( (d_1 - d_2(1,3))
+ 2 m_{\tau}^2 d_1/E_1E_3\right)/E_2E_4  \right]
\right\} \nonumber.
\end{eqnarray}

In these equations the energy  $\delta$-function can be trivially integrated 
away and we are left with  two-dimensional integral over the energies of 
incoming particles:
\begin{equation}
Hx\partial_x f_{\nu_i} ^{(1)} = \frac{G_F^2}{2\pi^3p_1} \int \int
A( F(f), d_k) \theta (E_3+E_4-E_1-m_2) E_2 p_3dp_3 
p_4dp_4,
\label{kin2a}
\end{equation}
where $E_2 = E'_1+E'_2-E_1 $,
$p_2=\sqrt{E_2^2-m_2^2}$, and
$A( F(f), d_k)$ is the integrand from the equations above.
$\theta (E_3+E_4-E_1-m_2)$ arises from the  energy conservation law.

\newpage
{\large \bf Figure Captions:}
\vskip1cm
\noindent
{\bf Fig. 1} $~~~$ 
Relative energy density of a massive tau neutrinos, 
$rm= \mnt \nnt /n_{\nu_0}$, for asymptotically large $x$  as a function of 
$\mnt$. The solid line corresponds to our results, while dashed, 
dashed-dotted and dotted lines correspond respectively to the results 
of refs.~\cite{ktcs,dr,hm}.

\vskip1cm
\noindent
{\bf Fig. 2} $~~~$ 
The quantity $T_{\gamma} a$ as a  function of the dimensionless
time $x$ for several values of the tau neutrino mass in the 
interval $0  < \mnt < 20 MeV$. The lowest and uppermost solid curves
correspond respectively to $\mnt = 0$ and to $\mnt = 20 MeV$, while 
the dashed curves correspond to several intermediate masses: 
$4,6,8,10,12,15 MeV$.

\vskip1cm
\noindent
{\bf Fig. 3} $~~~$ 
Relative deviation of the energy densities of massless neutrinos
from the equilibrium value, $\delta \rho/\rho = (\rho - \rho_{eq})/\rho_{eq}$, 
where $\rho_{eq} (x) =(7/120 \pi^2) (1 MeV/x)^4$. 
The solid and dashed lines correspond to $\nu_e$ and $\nu_\mu$ respectively.
In fig.~3a the evolution  of $\delta \rho/\rho$ with time $x$ 
for several different masses of $\nu_\tau$ is presented. 
In fig.~3b the asymptotic value of  $\delta \rho/\rho$ as a function of $\mnt$
is presented.

\vskip1cm
\noindent
{\bf Fig. 4} $~~~$ The relative deviation of the average energy
$(\langle E \rangle -\langle E_{eq}\rangle )/\langle E_{eq}\rangle $ for 
electronic and muonic neutrinos as a function of the mass of  tau neutrino. 
The solid line corresponds to electron neutrino, while the dashed one 
corresponds to muon neutrino.

\vskip1cm
\noindent
{\bf Fig. 5} $~~~$ 
Spectral distribution of electron neutrinos as 
function of the dimensionless momentum $y$.
Dotted, dashed, and solid lines correspond respectively to equilibrium 
distributions (initial condition), 
$\mnt = 20 MeV$ and $\mnt = 7 MeV$ at asymptotically large time.

\vskip1cm
\noindent
{\bf Fig. 6} $~~~$ 
The change in the electron neutrino 
spectral distribution times $y^2$ for various masses 
as a function of momentum $y$.
This quantity is related to the change in the neutron to proton
ratio, and therefore to the light element abundances.

\vskip1cm
\noindent
{\bf Fig. 7} $~~~$ 
Relative number of equivalent massless neutrino species $\Delta N = 
N_{eff} - 3$ as a function of $\nt$ mass, found from $^4 He$. 
Our result is presented by the 
solid line. Dashed, dashed-dotted, and dotted lines 
correspond respectively to the results  of refs.~\cite{hm,fko,ktcs}.

\vskip1cm
\noindent
{\bf Fig. 8} $~~~$
The number of  equivalent massless neutrino species $\Delta N = N_{eff} 
- 3$ as functions of the tau-neutrino mass, calculated from abundances 
of deuterium (solid), $^7Li$ (long dashed), $^3 He$ (dashed) and $^4 He$ 
(dotted).

\newpage
\psfig{file=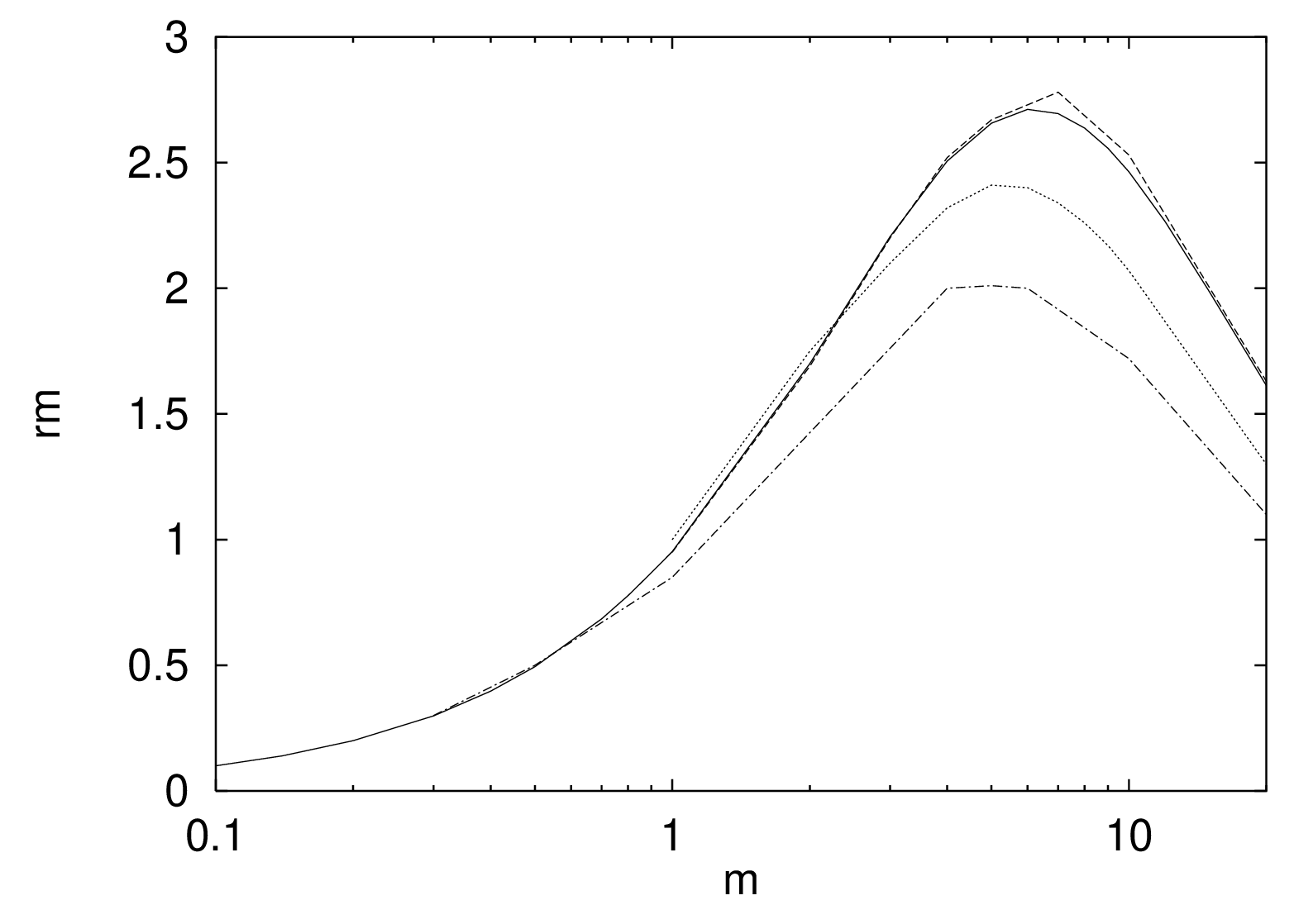,width=5in,height=3.5in}
\begin{center}
{\bf Figure 1.}
\end{center}

\psfig{file=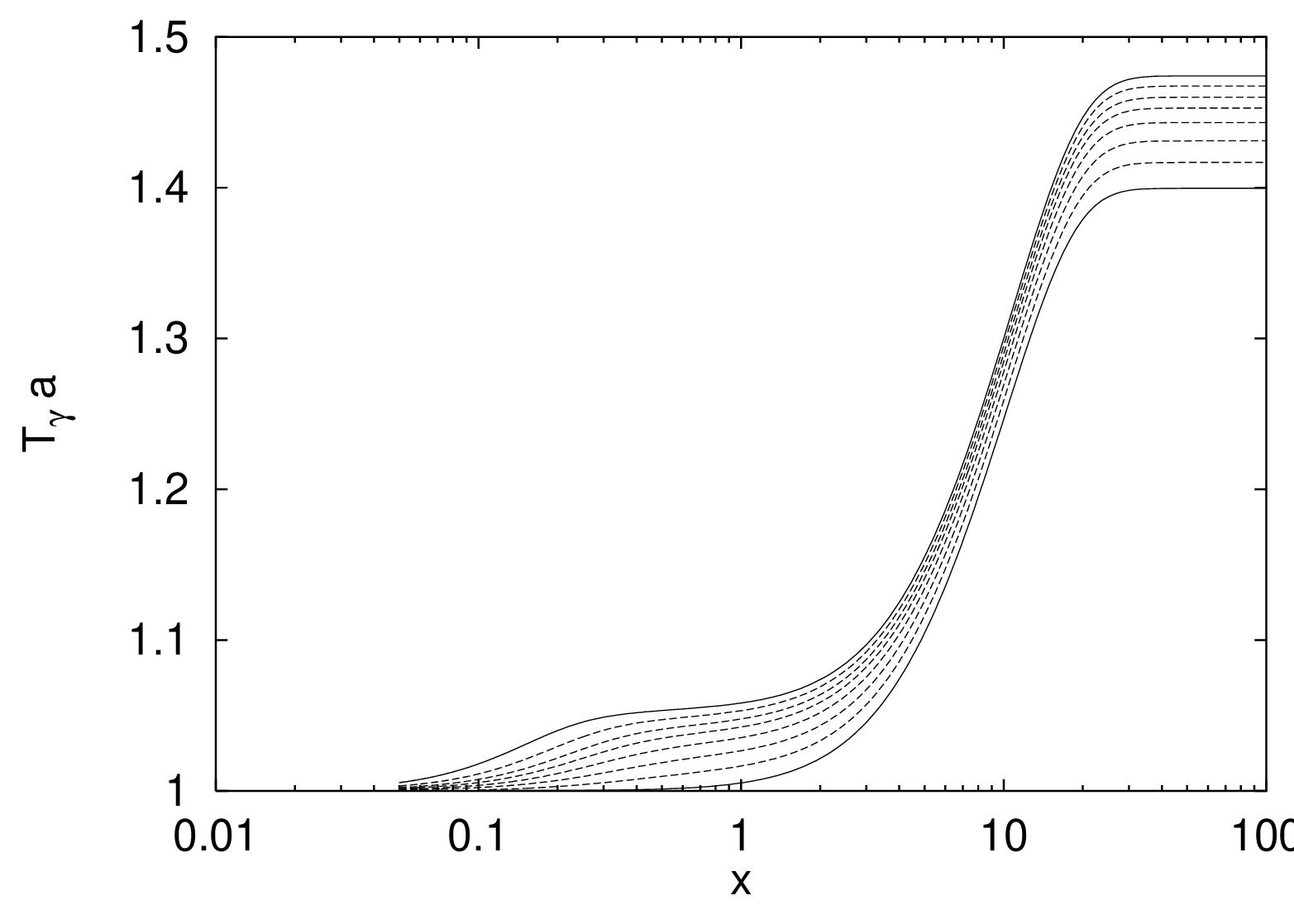,width=5in,height=3.5in}
\begin{center}
{\bf Figure 2.}
\end{center}

\newpage
\psfig{file=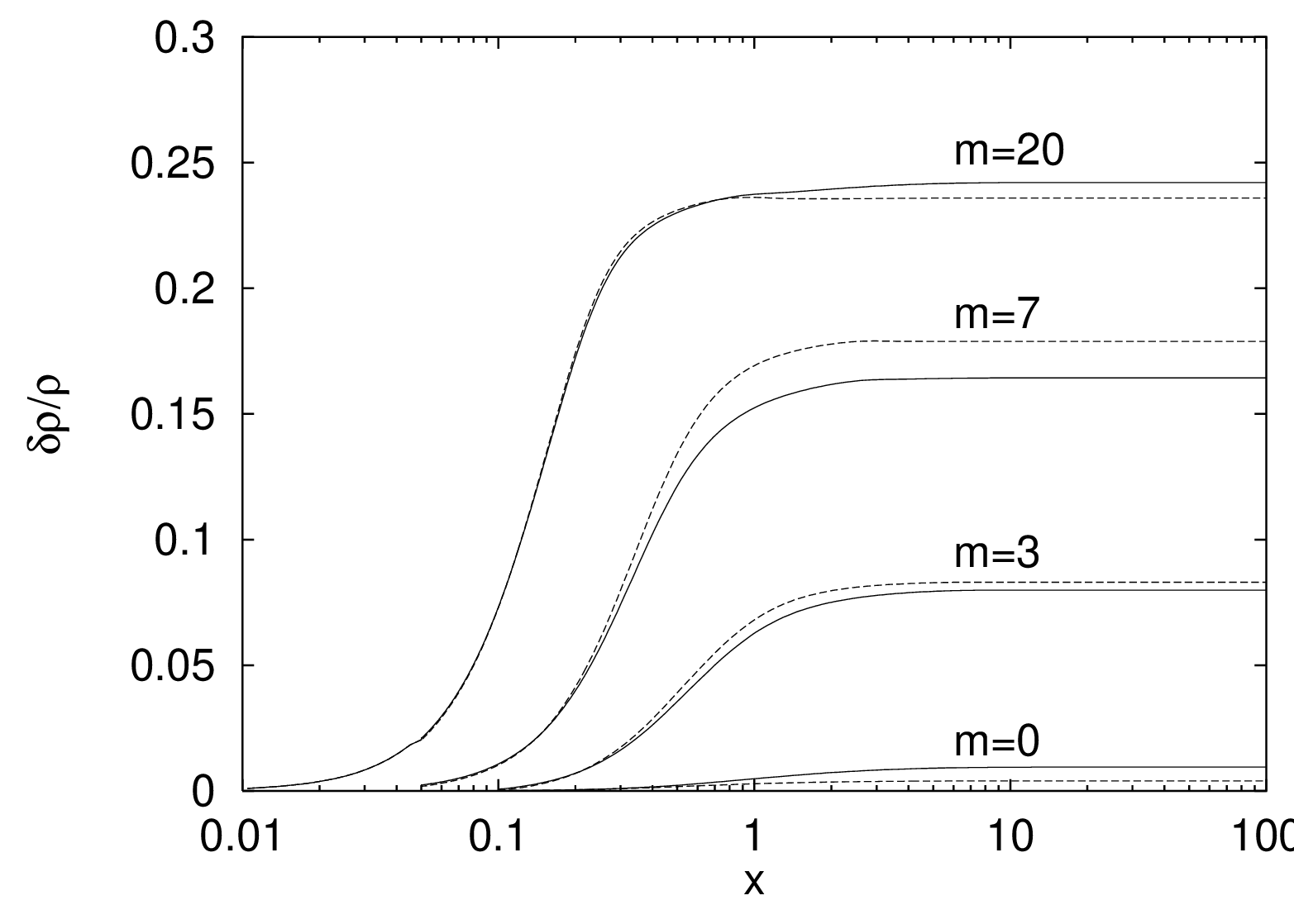,width=5in,height=3.5in}
\begin{center}
{\bf Figure 3a.}
\end{center}

\psfig{file=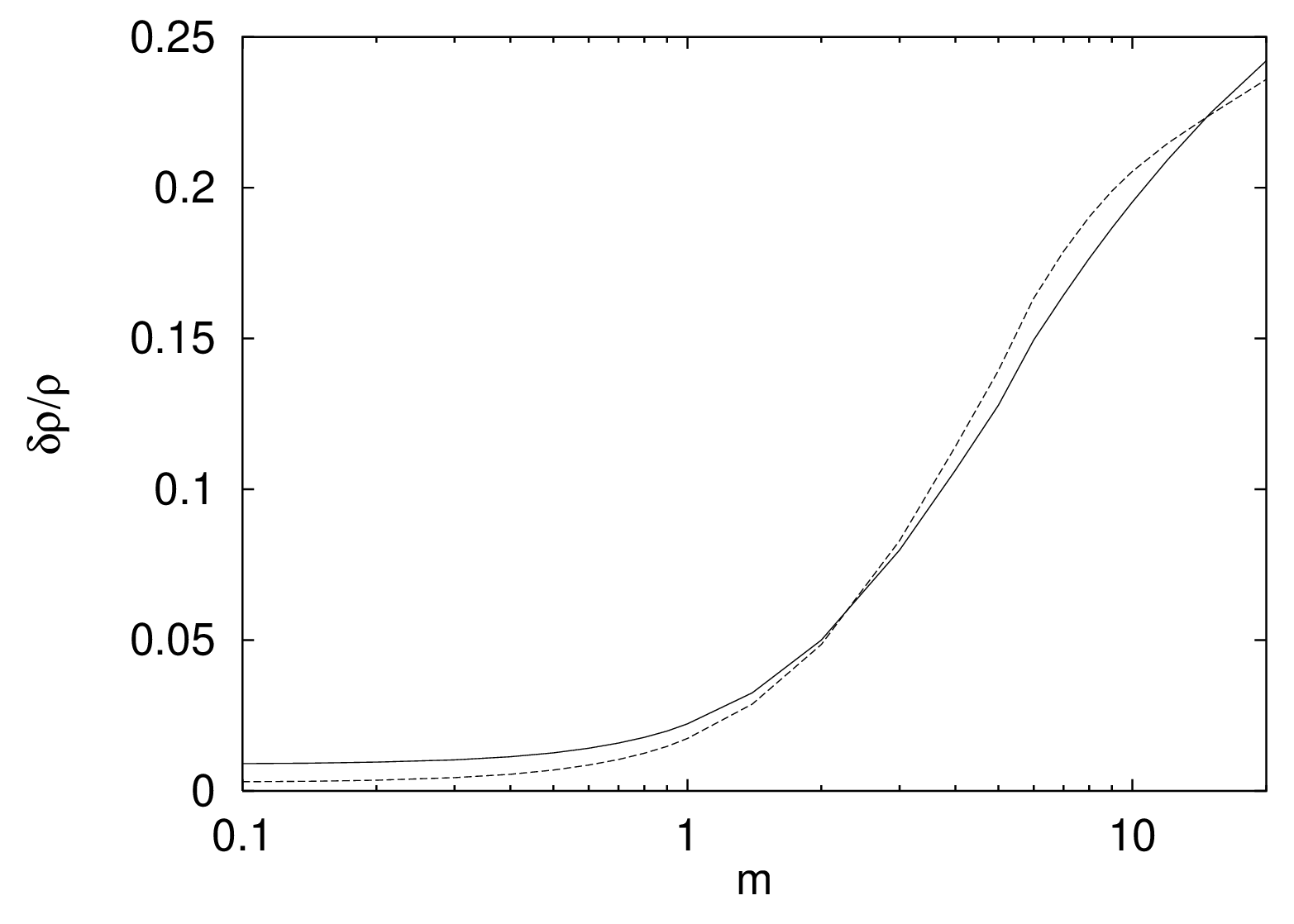,width=5in,height=3.5in}
\begin{center}
{\bf Figure 3b.}
\end{center}

\newpage
\psfig{file=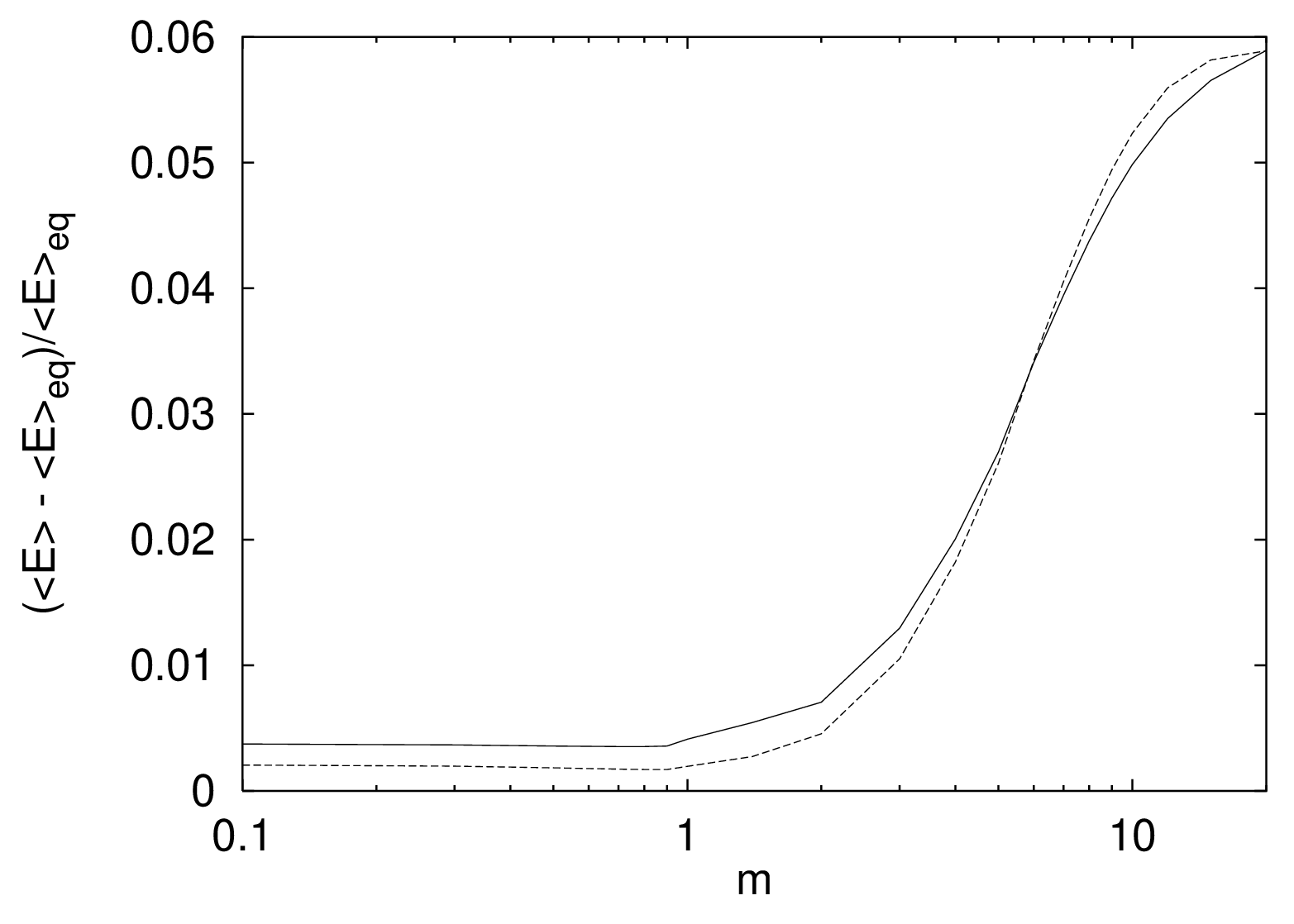,width=5in,height=3.5in}
\begin{center}
{\bf Figure 4.}
\end{center}

\psfig{file=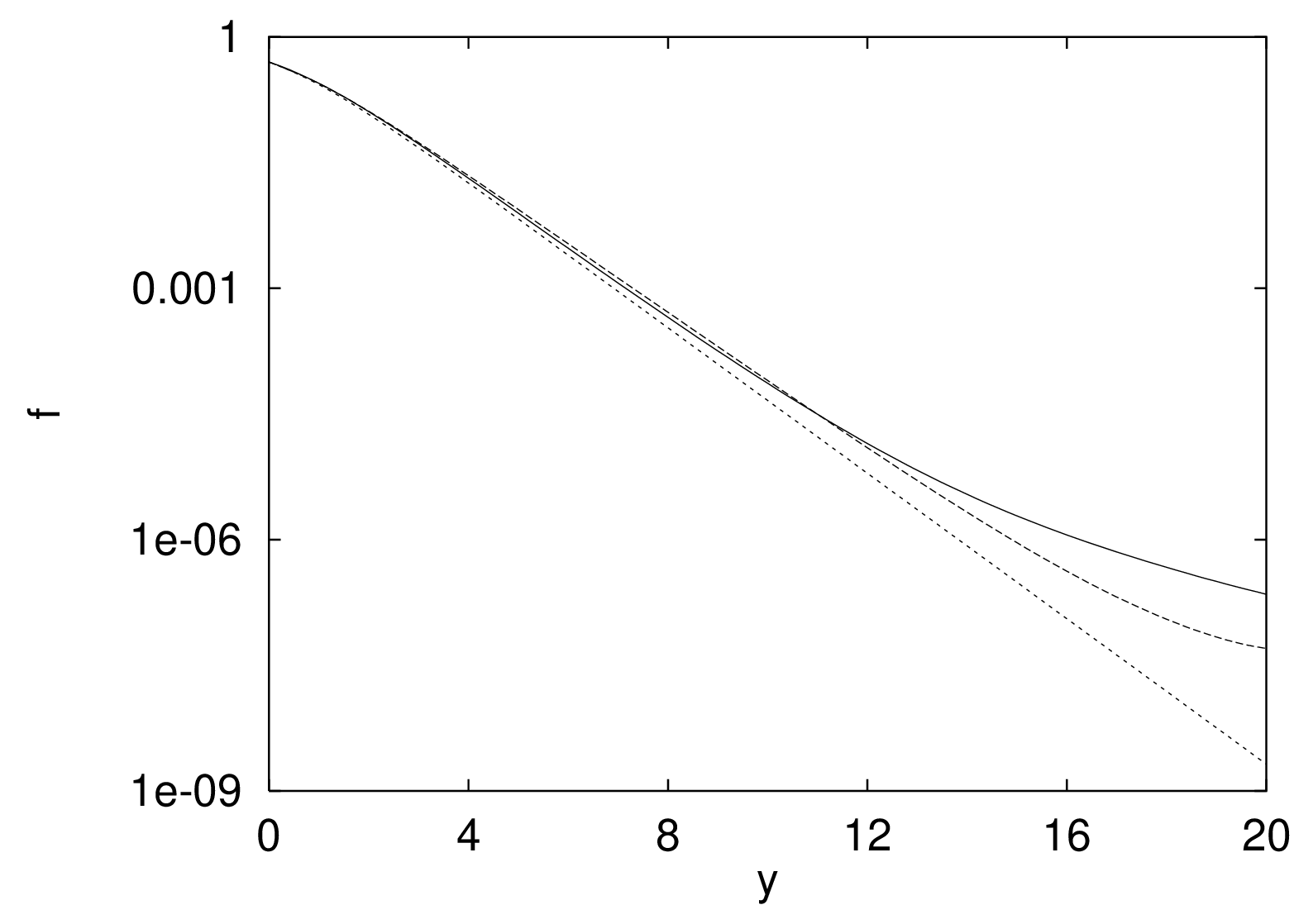,width=5in,height=3.5in}
\begin{center}
{\bf Figure 5.}
\end{center}

\newpage
\psfig{file=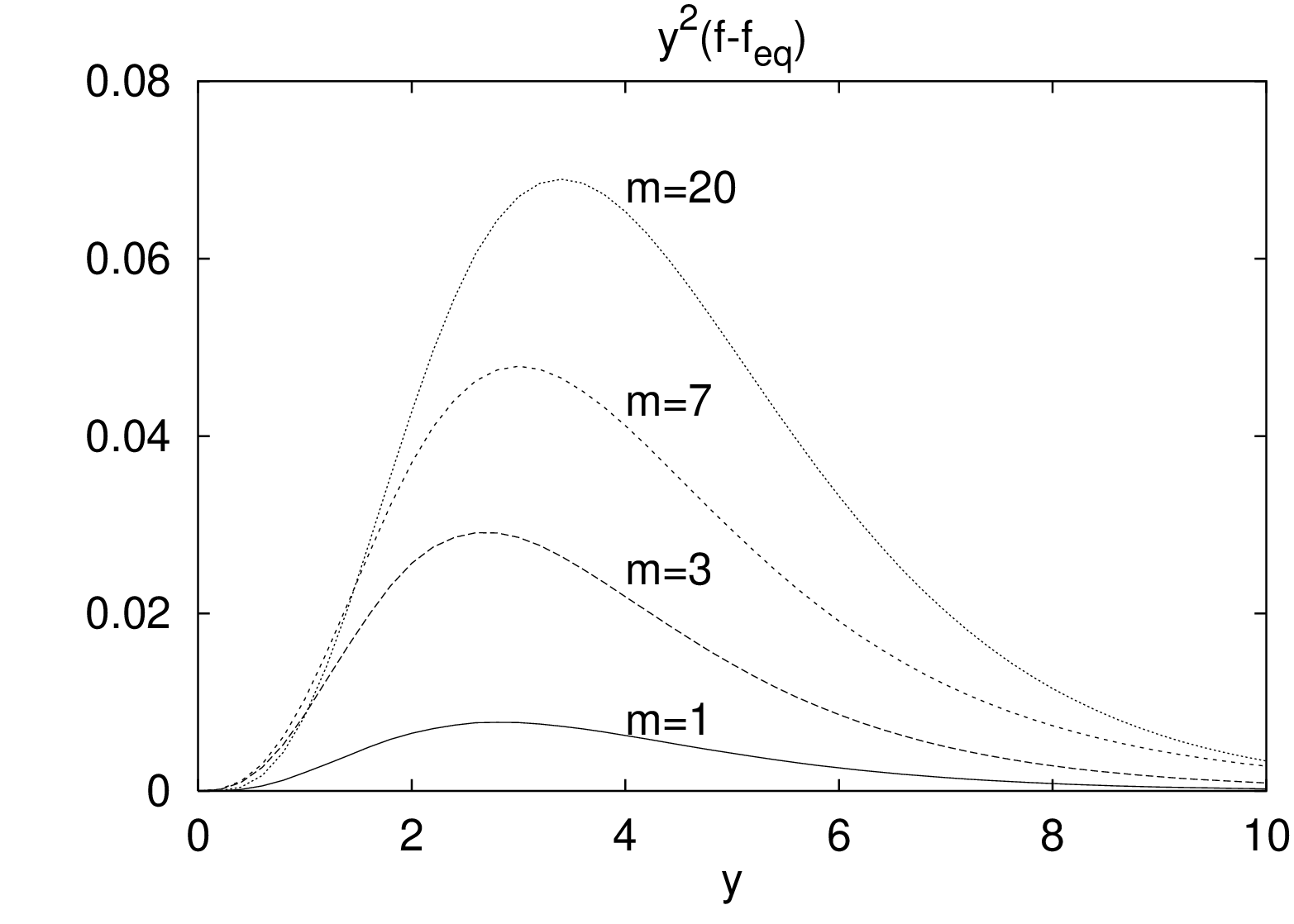,width=5in,height=3.5in}
\begin{center}
{\bf Figure 6.}
\end{center}

\psfig{file=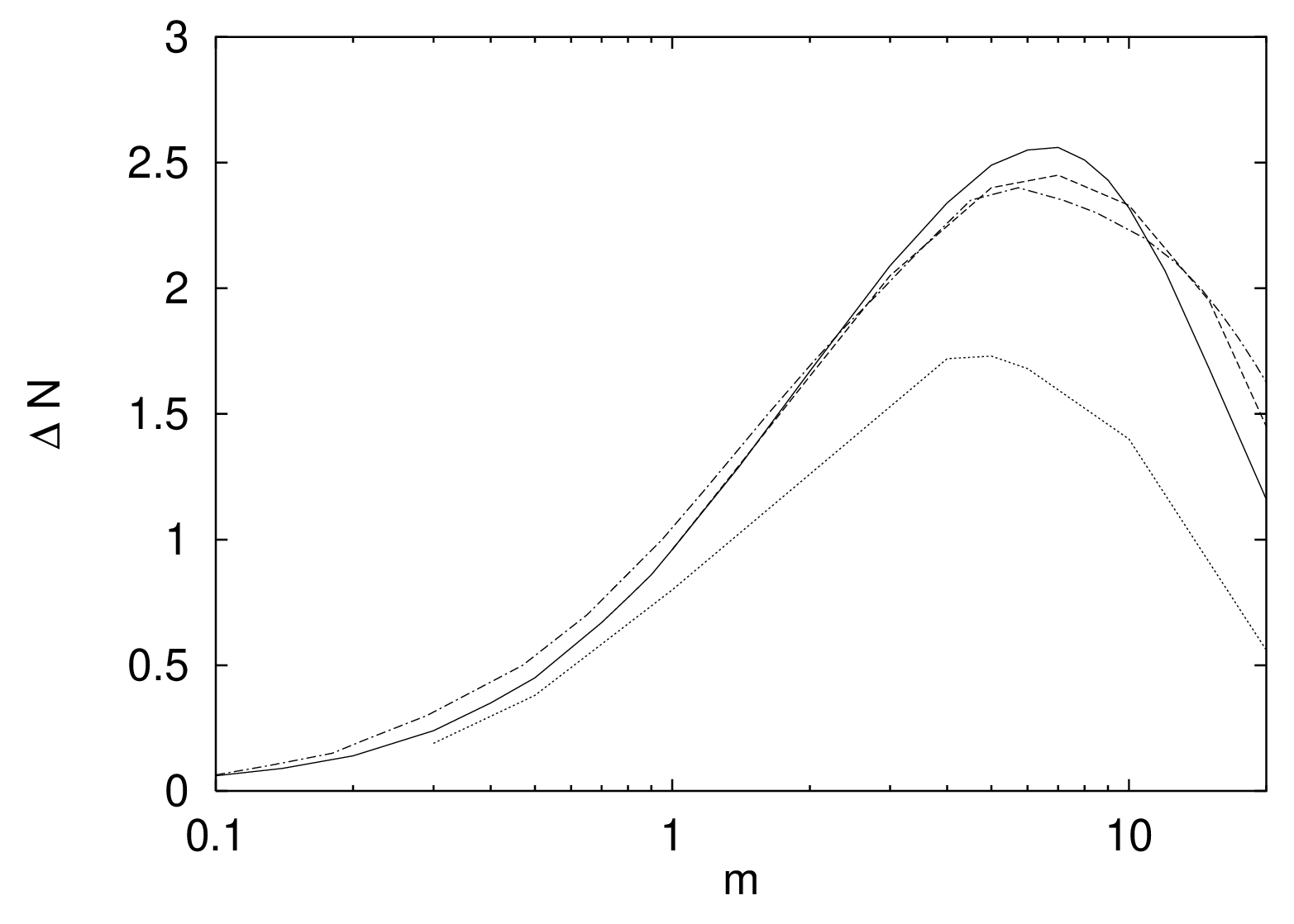,width=5in,height=3.5in}
\begin{center}
{\bf Figure 7.}
\end{center}
\newpage

\psfig{file=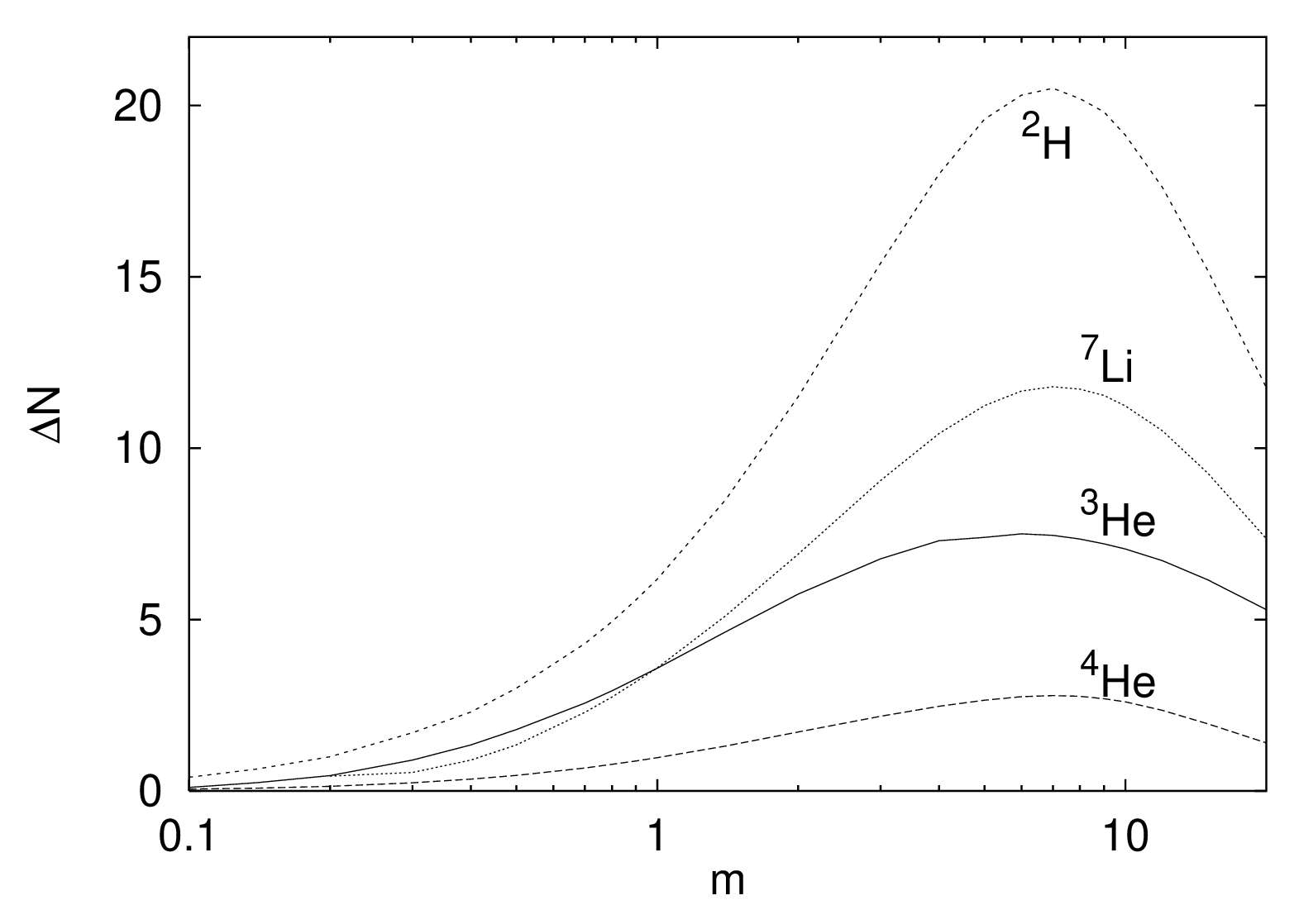,width=5in,height=3.5in}
\begin{center}
{\bf Figure 8.}
\end{center}

\newpage

\begin{table}
\begin{tabular}{|cccc|c|c|} 
 \hline\hline
&&&&&\\
& {\bf Process }&&&{\bf S }& {\bf $2^{-5} G^{-2}_{F} S \left| A \right| ^{2}$}\\
&&&&&\\
 \hline\hline
&&&&&\\
$\nu _{e} + \nu_{e} $ & $\rightarrow$ & $  \nu _{e} + \nu_{e} $&&1/4&
$2 \left[ (p_{1} \cdot p_{4}) (p_{2} \cdot p_{3}) 
 + (p_{1} \cdot p_{3}) (p_{2} \cdot p_{4}) \right.$\\
&&&&& $\left. ~~~~~~~~~~~~~~~~~~~~~~
 +  (p_{1} \cdot p_{2}) (p_{3} \cdot p_{4}) \right]$\\
&&&&&\\
\hline
&&&&&\\
$\nu _{e}+ \nu_{e} $ & $\rightarrow$ &
$ \nu _{\mu} + \nu_{\mu} $&&1/4&
$ \frac{1}{2} \left[ (p_{1} \cdot p_{4}) (p_{2} \cdot p_{3}) + 
(p_{1} \cdot p_{3}) (p_{2} \cdot p_{4}) \right]$\\
&&&&&\\
\hline
&&&&&\\
$\nu _{e}+ \nu_{e} $ & $\rightarrow$ &
$ \nu _{\tau} + \nu_{\tau} $&&1/4&
$ \frac{1}{2} \left[ (p_{1} \cdot p_{4}) (p_{2} \cdot p_{3}) +
(p_{1} \cdot p_{3}) (p_{2} \cdot p_{4}) \right.$\\
&&&&& - $\left. \mnt^2 (p_{1} \cdot p_{2}) \right]$\\
\hline
&&&&&\\
$\nu_{e} + \nu_\mu $ & $\rightarrow$&
$\nu _{e} + \nu_\mu  $&&1/2&
$ (p_{1} \cdot p_{2}) (p_{3} \cdot p_{4})
 + (p_{1} \cdot p_{4}) (p_{2} \cdot p_{3}) $\\
&&&&&\\
\hline
&&&&&\\
$\nu_{e} + \nu_\tau $ & $\rightarrow$&
$\nu _{e} + \nu_\tau  $&&1/2&
$  (p_{1} \cdot p_{2}) (p_{3} \cdot p_{4})
 + (p_{1} \cdot p_4) (p_{2} \cdot p_{3})  $\\
&&&&& 
 $  +  \mnt^2 (p_{1} \cdot p_{3}) $\\
&&&&&\\
\hline
&&&&&\\
$\nu _{e} + \nu_{e} $ & $\rightarrow$ & $  e^{+} + e^{-}$ &&1/2&
$2 ( g_{L}^2 + g_R^2) \left\{ (p_{1} \cdot p_{4}) (p_{2} \cdot p_{3}) 
\right.$ 
\\&&&&&$ \left.+  (p_{1} \cdot p_{3}) (p_{2} \cdot p_{4}) \right\}
+ 4 g_{L} g_{R} m _{e}^{2} (p_{1} \cdot p_{2})$\\
&&&&&\\
\hline
&&&&&\\
$\nu _{e} +  e^{\pm} $ & $\rightarrow$ & $ \nu _{e} +  e^{\pm} $&&1/2&
$ 2 (g_L^2 +g_R^2) \left\{ (p_{1} \cdot p_{2}) (p_{3} \cdot p_{4}) \right.$ \\
&&&&&$ \left. + (p_{1} \cdot p_{4}) (p_{2} \cdot p_{3}) \right\}
 - 4 g_{L} g_{R} m _{e}^{2} (p_{1} \cdot p_{3}) $\\
&&&&&\\
\hline
\hline
\end{tabular}
\end{table}
{\bf Table 1: } Matrix elements for various electron neutrino processes;
 $g_{L} = \frac{1}{2} + \sin^{2}\theta_{W}$ and
$g_{R} = \sin^{2} \theta_{W}$. Matrix elements for muon neutrino processes
are obtained by the substitutions $\nu_e \rightarrow \nu_\mu$ and  
$g_{L} \rightarrow \tilde{g}_{L} =g_{L} - 1$.

\newpage
\begin{table}
\begin{tabular}{|cccc|c|c|} 
 \hline\hline
&&&&&\\
& {\bf Process }&&&{\bf S }& {\bf $2^{-5} G^{-2}_{F} S \left| A \right| ^{2}$}\\
&&&&&\\
 \hline\hline
&&&&&\\
$\nu _\tau + \nu_\tau $ & $\rightarrow$ & $  \nu _\tau + \nu_\tau $&&1/4&
$2 \left[ (p_{1} \cdot p_{4}) (p_{2} \cdot p_{3}) 
 + (p_{1} \cdot p_{3}) (p_{2} \cdot p_{4}) \right.$\\
&&&&& $ 
 +  (p_{1} \cdot p_{2}) (p_{3} \cdot p_{4})  + 3\mnt^4 $\\
&&&&& $ 
 \left. +  2 \mnt^2 \left\{ (p_{1} \cdot p_{3}) + (p_{1} \cdot p_{4}) - 
(p_{1} \cdot p_{2}) \right\}   \right] $\\
\hline
&&&&&\\
$\nu _\tau+ \nu_\tau $ & $\rightarrow$ &
$ \nu _{e } + \nu_{e } $&&1/4&
$ \frac{1}{2} \left[ (p_{1} \cdot p_{4}) (p_{2} \cdot p_{3}) +
(p_{1} \cdot p_{3}) (p_{2} \cdot p_{4}) \right.$\\
&&&&& - $ \left. \mnt^2 (p_{3} \cdot p_{4}) \right]$\\
\hline
&&&&&\\
$\nu _\tau+ \nu_\tau $ & $\rightarrow$ &
$ \nu _\mu + \nu_\mu $&&1/4&
$ \frac{1}{2} \left[ (p_{1} \cdot p_{4}) (p_{2} \cdot p_{3}) +
(p_{1} \cdot p_{3}) (p_{2} \cdot p_{4}) \right.$\\
&&&&& - $ \left. \mnt^2 (p_{3} \cdot p_{4}) \right]$\\
\hline
&&&&&\\
$\nu_\tau + \nu_e  $ & $\rightarrow$&
$\nu _\tau + \nu_e   $&&1/2&
$ (p_{1} \cdot p_{2}) (p_{3} \cdot p_{4})
 + (p_{1} \cdot p_{4}) (p_{2} \cdot p_{3})  $\\
&&&&& 
 $  +  \mnt^2 (p_{2} \cdot p_{4}) $\\
\hline
&&&&&\\
$\nu_\tau + \nu_\mu $ & $\rightarrow$&
$\nu _\tau + \nu_\mu  $&&1/2&
$ (p_{1} \cdot p_{2}) (p_{3} \cdot p_{4})
 + (p_{1} \cdot p_{4}) (p_{2} \cdot p_{3})  $\\
&&&&& 
 $  +  \mnt^2 (p_{2} \cdot p_{4}) $\\
\hline
&&&&&\\
$\nu _\tau + \nu_\tau $ & $\rightarrow$ & $  e^{+} + e^{-}$ &&1/2&
$2 ( \tilde{g}_{L}^2 + g_R^2) \left\{ (p_{1} \cdot p_{4}) (p_{2} \cdot p_{3}) 
\right.$ 
\\&&&&&$ \left.+  (p_{1} \cdot p_{3}) (p_{2} \cdot p_{4}) 
- \mnt^2 (p_{3} \cdot p_{4}) \right\}$
\\&&&&& $ + 4 \tilde{g}_{L} g_{R} m _{e}^{2} \left\{ (p_{1} \cdot p_{2})
-2 \mnt^2 \right\}$\\
\hline
&&&&&\\
$\nu _\tau +  e^{\pm} $ & $\rightarrow$ & $ \nu _\tau +  e^{\pm} $&&1/2&
$2  (\tilde{g}_L^2 +g_R^2) \left\{ (p_{1} \cdot p_{2}) (p_{3} \cdot p_{4}) \right.$ \\
&&&&&$ \left. + (p_{1} \cdot p_{4}) (p_{2} \cdot p_{3}) 
 + \mnt^2 (p_{2} \cdot p_{4}) \right\}$
\\&&&&& $ - 4 \tilde{g}_{L} g_{R} m _{e}^{2} \left\{ (p_{1} \cdot p_{3}) 
+ 2 \mnt^2 \right\}$\\
\hline
\hline
\end{tabular}
\end{table}
{\bf Table 2: } Matrix elements for various tau-neutrino processes;
 $\tilde{g}_{L} =g_{L} - 1 = 
- \frac{1}{2} + \sin^{2}\theta_{W}$ and
$g_{R} = \sin^{2} \theta_{W}$.

\end{document}